\documentclass[letterpaper,twocolumn,amsmath,amssymb,jcp,10pt,aip]{revtex4-1}
\usepackage{graphicx}
\usepackage{dcolumn}
\usepackage{bm}
\usepackage{color}

\newcommand{\rr}{\textbf{r}}

\newcommand{\derivation}[1]{} 

\begin{document}
\title{Using Fundamental Measure Theory to Treat the Correlation
  Function of the Inhomogeneous Hard-Sphere Fluid}

\author{Jeff Schulte}
\author{Patrick Kreitzberg}
\author{Chris Haglund}
\author{David Roundy}
\affiliation{Department of Physics, Oregon State University, Corvallis, OR 97331}

\begin{abstract}
  We investigate the value of the correlation function of an
  inhomogeneous hard-sphere fluid at contact.  This
  quantity plays a critical role in Statistical Associating Fluid
  Theory (SAFT), which is the basis of a number of recently developed classical
  density functionals.  We define two averaged values for the
  correlation function at contact, and derive formulas for each of
  them from the White Bear version of the Fundamental Measure Theory
  functional~\cite{roth2002whitebear}, using an assumption of
  thermodynamic consistency. We test these formulas, as well as two
  existing formulas~\cite{yu2002fmt-dft-inhomogeneous-associating,
    gross2009density} against Monte Carlo simulations, and find
  excellent agreement between the Monte Carlo data and one of our
  averaged correlation functions.
\end{abstract}

\maketitle

\section{Introduction}

\newcommand\saftlocaldft{felipe2001examination, gloor2002saft,%
  gloor2004accurate, clark2006developing, gloor2007prediction,%
  kahl2008modified, gross2009density}
\newcommand\saftnonlocaldft{yu2002fmt-dft-inhomogeneous-associating,%
  fu2005vapor-liquid-dft,bryk2006density}

There has been considerable recent interest in using Statistical
Associating Fluid Theory (SAFT) to construct classical density
functionals to describe associating
fluids\cite{\saftlocaldft,\saftnonlocaldft}.  This approach has been
successful in qualitatively describing the dependence of surface
tension on temperature.
%
%
A key input to these functionals is the correlation function
evaluated at contact, which is required for both the chain and
association terms in the SAFT free energy.  The chain term describes
the chain formation energy in polymeric fluids, while the association
term describes the effects of hydrogen bonding, both of which can be
large.
Yu and Wu introduced in 2002 a functional for the association term of
the free energy, which included a functional for the contact value of
the correlation function (described in
Section~\ref{sec:yuwu})\cite{yu2002fmt-dft-inhomogeneous-associating},
which has subsequently been used in the development of other
SAFT-based functionals\cite{fu2005vapor-liquid-dft, bryk2006density}.
Two functionals for the chain contribution have recently been
introduced, one which uses the
correlation function of Yu and Wu\cite{bryk2006density} and another that
introduces a new approximation for the contact value of the
correlation function (described in
Section~\ref{sec:gross})\cite{gross2009density}.
Given these different approaches, it seems valuable to examine this
property of the hard-sphere fluid through direct simulation, in order
to establish the advantages and disadvantages of each approach.

Although these recent works have introduced approximate functionals
for the contact value of the correlation
function\cite{yu2002fmt-dft-inhomogeneous-associating,
  gross2009density}, there has not been a study that specifically
addresses this contact value for an inhomogeneous hard-sphere fluid.
In this paper we introduce two definitions for the locally averaged
correlation function of an inhomogeneous system.
%
%
Given these definitions, we will present a thermodynamic derivation
for each correlation function from the free energy functional.  We
will then discuss the correlation functions of Yu and Wu and of Gross, and
will end by comparing all four approximations with Monte-Carlo
simulations of the hard-sphere fluid at a variety of hard-wall
surfaces.

\section{Correlation function with inhomogeneity}

We define our terms using the two-particle density
$n^{(2)}(\rr_1,\rr_2)$, which gives the probability per unit volume
squared of finding one particle at position $\rr_1$ and the other at
position $\rr_2$.  The pair correlation function is defined by
\begin{align}
  g(\rr_1,\rr_2) &\equiv \frac{n^{(2)}(\rr_1,\rr_2)}{n(\rr_1)n(\rr_2)}
\end{align}
In a homogeneous fluid, the pair correlation only depends on the
distance $|\rr_1-\rr_2|$ and can be expressed as a function of a
single variable. The contact value of the correlation function is this
correlation function's value when evaluated at a distance of the
diameter $\sigma$.  It is desirable for reasons of efficiency to limit CDFT
functionals to one-center convolutions, which leads us to seek a
simplified expression for the contact value of the correlation
function---which is the same as the contact value of the cavity
correlation function for hard spheres.
In a system with an inhomogeneous density, we seek a \emph{local}
value for $g_\sigma$.  There are two reasonable options for defining
such a local function: a symmetric formulation (which we refer to as $S$) and an
asymmetric formulation (which we refer to as $A$).

\begin{figure}
\includegraphics[width=5cm]{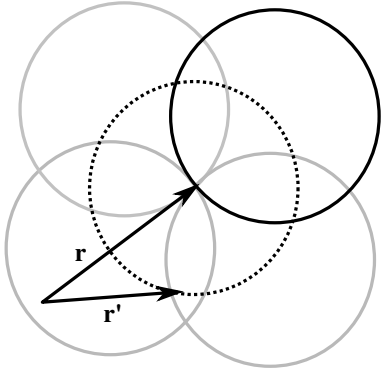}
\caption{Set of hard spheres that are included in $n_0(\mathbf{r})$, which
  consist of those which just touch the point $\mathbf{r}$.}
\label{fig:n0}
\end{figure}

For the symmetric $S$ case, the correlation function at contact is
given by:
\begin{align}
  g^S_\sigma(\rr) &= \frac{1}{n_0(\rr)^2}\int n^{(2)}(\rr - \rr', \rr
  + \rr')
  \frac{\delta(\sigma/2 -|\rr'|)}{\pi\sigma^2}d\rr' \label{eq:gS}
\end{align}
where $\sigma$ is the hard sphere diameter and density $n_0$ is one of the fundamental measures of Fundamental
Measure Theory (FMT), and is ideal for treating touching spheres, as
illustrated in Figure~\ref{fig:n0}.
\begin{align}
  n_0(\rr) &= \int n(\rr')\frac{\delta(\sigma/2 -|\rr-\rr'|)}{\pi\sigma^2} d\rr'
\end{align}
This functional gives a value averaged over all spheres that touch at
the position $\rr$.

In contrast, the asymmetrically averaged $A$ correlation function is
given by
\begin{align}
  g^A_\sigma(\rr) &= \frac{1}{n(\rr)n_A(\rr)}
  \int n^{(2)}(\rr, \rr + \rr')
  \frac{\delta(\sigma -|\rr'|)}{4\pi\sigma^2}d\rr' \label{eq:gA}
\end{align}
where the density $n_A(\rr)$ is similar to $n_0$, but measures the
density of spheres that are touching a sphere that is located at
point $\rr$.
\begin{align}
  n_A(\rr) &= \int n(\rr')
  \frac{\delta(\sigma -|\rr-\rr'|)}{4\pi\sigma^2} d\rr' \label{eq:nA}
\end{align}
Thus $g_\sigma^A$ corresponds to an average of the two-particle
density over spheres touching a sphere that is located at the
position~$\rr$.

\newcommand\ncontact{\ensuremath{n_\textit{contact}}}

\subsection*{Fundamental-Measure Theory}

We use the White Bear version of the Fundamental-Measure Theory~(FMT)
functional~\cite{roth2002whitebear}, which describes the excess free
energy of a hard-sphere fluid.  The White Bear functional reduces to
the Carnahan-Starling equation of state for homogeneous systems.  It
is written as an integral over all space of a local function of a set
of ``fundamental measures'' $n_\alpha(\rr)$, each of which is written
as a one-center convolution of the density.  The White Bear free
energy is thus
\begin{equation}
A_\textit{HS}[n] = k_B T \int \left(\Phi_1(\rr) + \Phi_2(\rr) + \Phi_3(\rr)\right) d\rr \; ,
\end{equation}
with integrands
\begin{align}
\Phi_1 &= -n_0 \ln\left( 1 - n_3\right) \label{eq:Phi1}\\
\Phi_2 &= \frac{n_1 n_2 - \mathbf{n}_{V1} \cdot\mathbf{n}_{V2}}{1-n_3} \\
\Phi_3 &= (n_2^3 - 3 n_2 \mathbf{n}_{V2} \cdot \mathbf{n}_{V2}) \frac{
  n_3 + (1-n_3)^2 \ln(1-n_3)
}{
  36\pi n_3^2\left( 1 - n_3 \right)^2
} , \label{eq:Phi3}
\end{align}
using the fundamental measures
\begin{align}
  n_3(\rr) &= \int n(\rr') \Theta(\sigma/2 -\left|\rr - \rr'\right|)
  d\rr' \label{eq:FMn3} \\
  n_2(\rr) &= \int n(\rr') \delta(\sigma/2 -\left|\rr - \rr'\right|) d\rr' \\
  \mathbf{n}_{2V}(\rr) &= \int n(\rr') \delta(\sigma/2 -\left|\rr - \rr'\right|) \frac{\rr-\rr'}{|\rr-\rr'|}d\rr'
\end{align}
\begin{align}
  \mathbf{n}_{V1} = \frac{\mathbf{n}_{V2}}{2\pi \sigma}, \quad
  n_1 &= \frac{n_2}{2\pi \sigma} , \quad
  n_0 = \frac{n_2}{\pi \sigma^2} \label{eq:FMrest}
\end{align}

\section{Theoretical Approaches}

\subsection{Homogeneous limit}

In order to motivate our derivation of the correlation function at
contact for the \emph{inhomogeneous} hard-sphere fluid, we begin by
deriving the well-known formula for $g_\sigma$ for the
\emph{homogeneous} fluid that comes from the Carnahan-Starling free
energy.  The contact value of the correlation function density can be
found by using the contact-value theorem, which states that the
pressure on any hard surface is determined by the density at contact:
\begin{align}
  p &= k_BT n_\textit{contact} \\
  &= k_BT n g_\sigma
\end{align}
Since we are interested in the correlation function at the surface of
the hard spheres, we need to compute the pressure on that surface.  The pressure on a hard sphere can be
readily computed from the dependence of the Carnahan-Starling free
energy on hard sphere radius.
\begin{align}
  A_{HS} &= Nk_BT \frac{4\eta - 3\eta^2}{(1-\eta)^2}
\end{align}
where $\eta \equiv \frac{\pi}{6} \sigma^3 n$ is the filling fraction.
We can thus readily compute the derivative of free energy with respect
to hard-sphere diameter, which is half the force on \emph{all} the
hard spheres---since we're changing all their diameters at once.  To
compute the pressure on the spheres, we need to divide the total force
by the total area of contact, which means dividing by $N 4\pi
\sigma^2$.
\begin{align}
  p_\sigma &= \frac{1}{N 4\pi \sigma^2} \frac12 \frac{dA_{HS}}{d\sigma} \\
  &= k_BT n \frac{1 - \frac{\eta}2}{(1-\eta)^3}
\end{align}
Using the contact-value theorem, we can thus find the well-known
correlation function evaluated at contact.
\begin{align}
  g_\sigma &= \frac{1 - \frac{\eta}2}{(1-\eta)^3} \label{eq:cs-g}
\end{align}
Extending this derivation to the inhomogeneous fluid requires that we
find the pressure felt by the surface of particular spheres.

\subsection{Asymmetrically averaged correlation function}\label{sec:g-A}

We will begin our derivation of the locally averaged correlation
function with the asymmetric definition of $g_\sigma^A(\rr)$ given in
Equation~\ref{eq:gA}, which is averaged over contacts in which one of
the two spheres is located at position~$\rr$.  This correlation
function is related to the contact density averaged over the surface
of a sphere located at~$\rr$, and can thus be determined by finding
the pressure on that sphere.  We find this pressure from the change in
free energy resulting from an infinitesimal expansion of any
spheres located at position~$\rr$.  From this pressure, we derive a
formula for the correlation function $g_\sigma^A(\rr)$ as was done in
the previous section:
\begin{align}
  p_\sigma(\mathbf{r}) &= \frac{1}{n(\rr) 2\pi \sigma^2} \frac{\delta
    A_{HS}}{\delta \sigma(\mathbf{r})} \label{eq:p_{HS}^A}\\
  g_\sigma^A(\rr)
  &= \frac{1}{n(\rr) n_A(\rr)}\frac{1}{ k_BT 2\pi \sigma^2} \frac{\delta
    A_{HS}}{\delta \sigma(\mathbf{r})} \label{eq:g-A-exact}
\end{align}
Details of evaluating the functional derivative in
Equation~\ref{eq:g-A-exact} using FMT are given in
the appendix.  The equation for $g_\sigma^A$ requires
finding convolutions of local derivatives of the free energy, making
this formulation computationally somewhat more expensive than the free
energy itself.

\derivation{
  \end{widetext}
}

\subsection{Symmetrically averaged correlation function}\label{sec:g-S}

We now address the symmetrically averaged correlation function, which
is defined in Equation~\ref{eq:gS}.  This corresponds to the
correlation function averaged for spheres \emph{touching at a given
  point}.  In this case, we conceptually would like to evaluate the
pressure felt by the surface of spheres where that surface is located
at point~$\rr$.  We can approximate this value by assuming that this
pressure will be simply related to the free energy density at
point~$\rr$.  Through a process similar to the previous derivations, this
leads to the expression
\begin{align}
  g_\sigma^S(\rr)
  &= \frac{1}{n_0(\rr)^2}\frac{1}{ 2\pi \sigma^2}
  \frac{\partial \Phi(\rr)}{\partial \sigma} \label{eq:g-S}
\end{align}
where $\Phi(\rr) = \Phi_1(\rr) + \Phi_2(\rr) + \Phi_3(\rr)$ is the
dimensionless free energy density.  This step is an
approximation---unlike the analogous
Equation~\ref{eq:g-A-exact}---because it assumes that we have
available a local functional $\Phi(\rr)$ whose derivative provides the
pressure needed to compute $g_\sigma(\rr)$.  Equation~\ref{eq:g-S}
requires that we evaluate the derivatives of the fundamental measures
$n_\alpha(\rr)$ with respect to diameter, which leads us to
derivatives of the $\delta$~function, which we can simplify and
approximate using an assumption of a reasonably smooth density:
\begin{align}
  \frac{\partial n_2(\rr)}{\partial \sigma}
  &= \frac12 \int \delta'(\frac\sigma2 - |\rr-\rr'|) n(\rr')d\rr' \\
  &= \frac2{\sigma}n_2(\rr) - \frac12 \int \delta\left(\frac\sigma2 - |\rr-\rr'|\right)
  \frac{\rr-\rr'}{|\rr-\rr'|}\cdot\mathbf{\nabla}n(\rr')d\rr' \\
  &\approx \frac{2}{\sigma}n_2(\rr)
\end{align}
In the systems that we study, the density is \emph{not} reasonably
smooth, but we can state empirically making this approximation
nevertheless improves the predictions of our functional, while at the
same time reducing its computational cost by avoiding the need to
calculate any additional weighted densities or convolutions.

\subsection{Gross's asymmetrically averaged correlation functional}\label{sec:gross}
One approximation for the correlation function is that of
Gross\cite{gross2009density}, which is of the asymmetrically averaged
variety ($g_\sigma^A$):
\begin{align}
  g_\sigma^\text{Gross,A}(\rr) &= \frac{1 - \frac{\pi}{12}\sigma^3n_A(\rr)}{\left(1 -
    \frac{\pi}{6}\sigma^3n_A(\rr)\right)^3}
\end{align}
where $n_A$ is the averaged density defined in Equation~\ref{eq:nA}.
This formula is arrived at by using the density averaged over all
spheres that could be touching a sphere at point~$\rr$ in the
Carnahan-Starling equation for the correlation function at contact,
given in Equation~\ref{eq:cs-g}.

\subsection{Yu and Wu's symmetrically averaged functional}\label{sec:yuwu}

Yu and Wu developed a functional for the correlation function
evaluated at contact which is symmetrically
averaged~\cite{yu2002fmt-dft-inhomogeneous-associating}.  However,
instead of using $n_0$ as the corresponding density, they use a
density given by
\begin{align}
  n_\text{Yu}(\rr) &= n_0(\rr) \zeta(\rr) \label{eq:nYu} \\
  \zeta &= 1 - \frac{\mathbf{n_2}\cdot\mathbf{n_2}}{n_2^2}
\end{align}
where the function $\zeta$ is a measure of local inhomogeneity at the
point of contact, and has the effect of reducing this density at
interfaces.  Because of this difference, the correlation function of
Yu and Wu cannot be directly compared with $g_\sigma^S$ as defined in
Equation~\ref{eq:gS}.  Therefore in order to make a comparison we move
the factors of $\zeta$ in Equation~\ref{eq:nYu} from the density into
the correlation function itself.
\begin{align}
  g_\sigma^\text{Yu,S} &= \zeta^2 g_\sigma^\text{Yu} \\
   &= \zeta^2\left(\frac{1}{1-n_3}
    + \frac14 \frac{\sigma n_2\zeta}{(1-n_3)^2}
    + \frac1{72} \frac{\sigma^2 n_2^2 \zeta}{(1-n_3)^3}\right)
\end{align}
where $g_\sigma^\text{Yu}$ is the correlation function as defined in
reference~\cite{yu2002fmt-dft-inhomogeneous-associating}, and
$g_\sigma^\text{Yu,S}$ is the function we will examine in this paper.



\begin{figure}
  \includegraphics[width=\columnwidth]{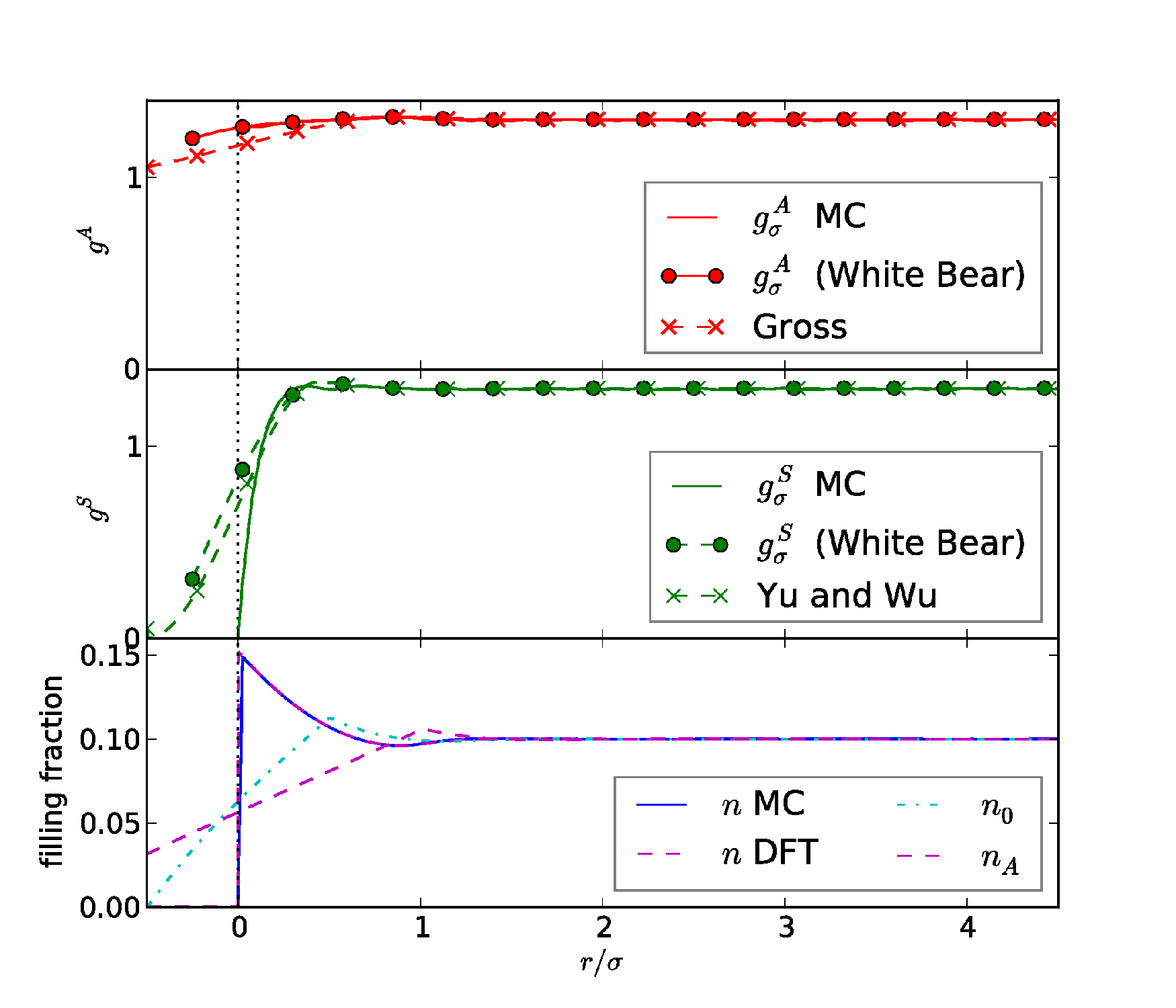}
  \caption{Density and correlation function at a hard wall with bulk
    filling fraction of 0.1.}
  \label{fig:walls-10}
\end{figure}

\section{Comparison with simulation}\label{sec:comparison}

We performed a Monte-Carlo simulation of the hard sphere fluid to
measure the contact value of the correlation function for several
simple inhomogeneous configurations.  For each configuration, we
compute the mean density, and the contact values of the correlation
function, averaged as defined in Equations~\ref{eq:gA} and
\ref{eq:gS}.  We compare these with the functionals presented in
sections~\ref{sec:g-A} to~\ref{sec:yuwu}.  We constructed our
functionals using both the original White Bear
functional~\cite{roth2002whitebear} as well as the mark II version of
the White Bear functional~\cite{hansen2006density}, but the results
were essentially indistinguishable on our plots, so we exclusively
show the results due to the original White Bear functional.

We simulate the inhomogeneous hard sphere fluid at four hard-wall
interfaces.  The first and simplest is a flat hard wall.  We
then study two convex hard surfaces.  One is an excluded sphere with
diameter $2\sigma$, which corresponds to a ``test particle''
simulation with one of a hard sphere at the origin with diameter
$\sigma$.  The second is an excluded sphere with diameter $6\sigma$,
which demonstrates behavior typical of mildly convex hard surfaces.
Finally, we study a concave surface given by a hard cavity in which
our fluid is free to move up to a diameter of $16\sigma$, which
demonstrates behavior typical of mildly concave surfaces.  In each
case, we performed a low-density (filling fraction 0.1) and high-density
(filling fraction 0.4) simulation.  We performed additional
computations over a wider range of curvatures and densities, but
chose these as typical examples.

\begin{figure}
  \includegraphics[width=\columnwidth]{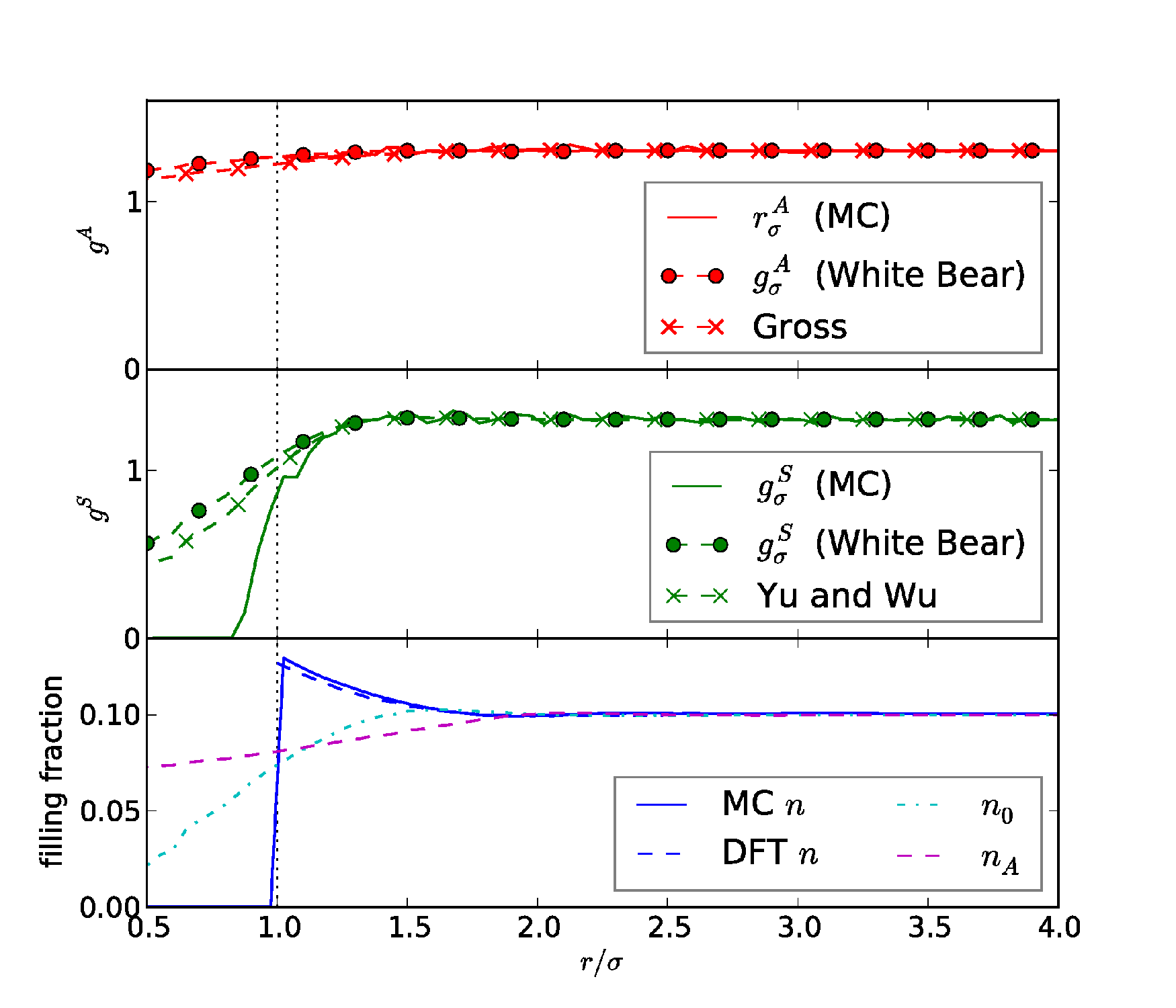}
  \caption{Density and correlation function around a hard sphere with
    diameter the same as those in the fluid, with a bulk filling
    fraction of 0.1.}
  \label{fig:inner-4-10}
\end{figure}

\subsection{Low density}

\begin{figure}
  \includegraphics[width=\columnwidth]{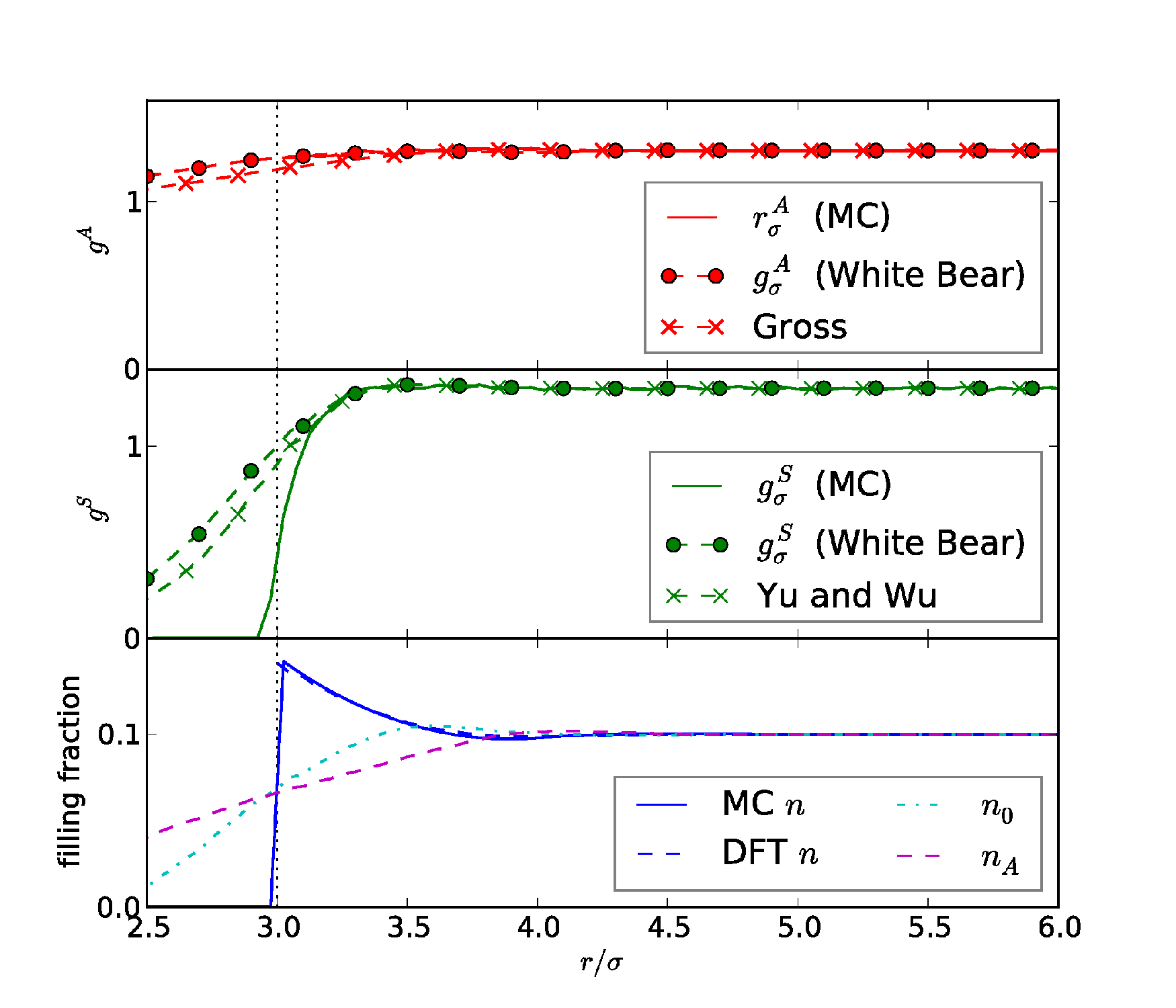}
  \caption{Density and correlation function around a hard sphere with
    an excluded diameter of $6\sigma$, with a bulk filling fraction of
    0.1.}
  \label{fig:inner-12-10}
\end{figure}

\begin{figure}
  \includegraphics[width=\columnwidth]{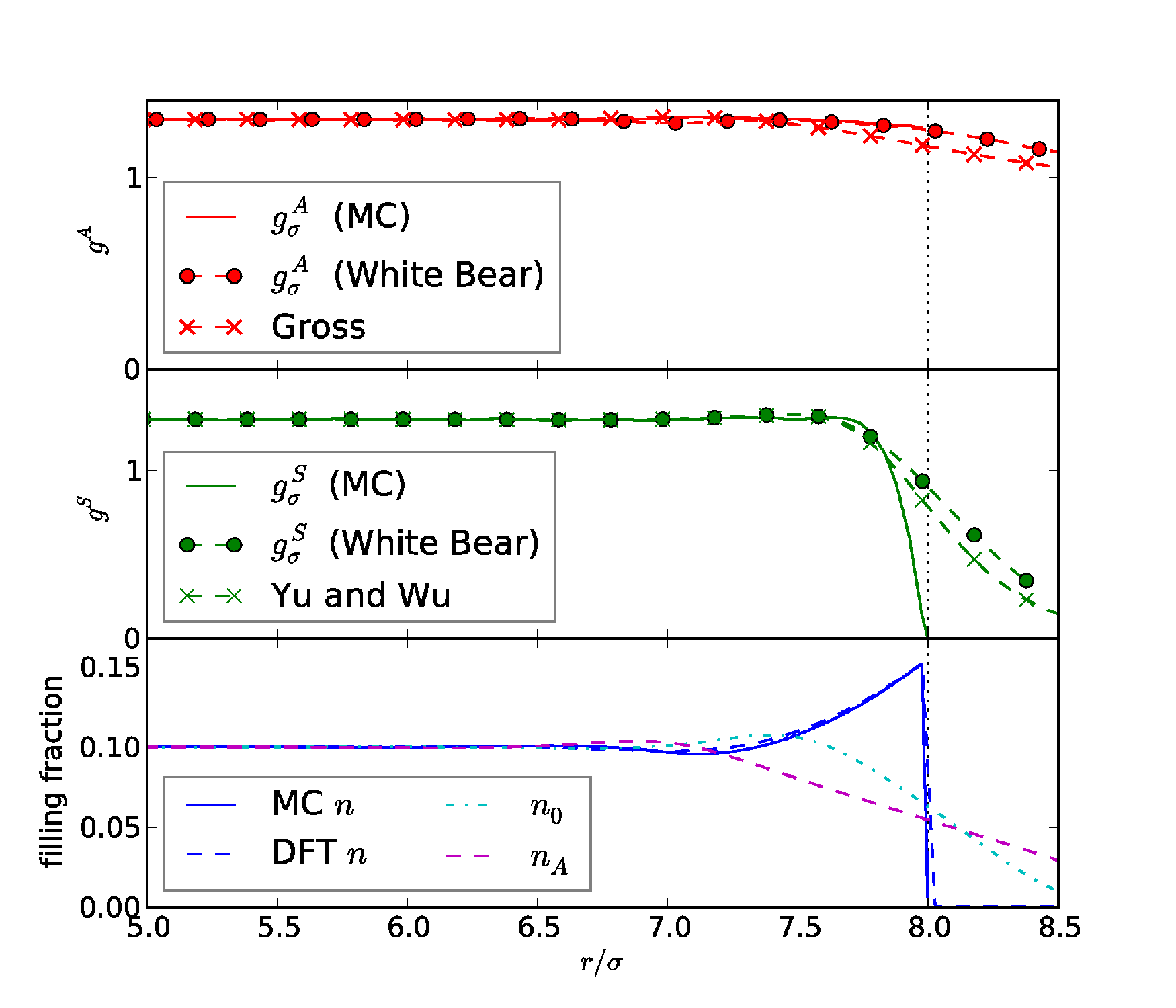}
  \caption{Density and correlation functions near the surface of a
    spherical cavity with diameter $16\sigma$ at a bulk filling
    fraction of 0.1.}
  \label{fig:outer-10}
\end{figure}

We begin by presenting our low-density results, corresponding to a
filling fraction of 0.1, which are shown in
Figures~\ref{fig:walls-10}-\ref{fig:outer-10}.  At this low density,
the contact value of the correlation function in the bulk is only 1.3,
indicating that correlations are indeed small and that the fluid should be
relatively easy to model.  Indeed, the contact density at the hard
surface is only around 50\% higher than the bulk, and the FMT
predicted density is close to indistinguishable from the true
density for each of the four configurations.

The $g_\sigma^A$ correlation function in each configuration is very
flat, with only small, smooth changes as the surface is approached.
Our functional $g_\sigma^A$ very closely matches the Monte Carlo
predictions in each case, while that of Gross consistently
underestimates the correlation at the interface by a significant margin.  We note that the theoretical curves
extend into the region from which the fluid is excluded.  This value
corresponds to the correlation function that would be observed in the
vanishingly unlikely scenario in which there was a sphere present at
that location.  Naturally, we are unable to observe this quantity in
our Monte Carlo simulations.

The $g_\sigma^S$ correlation function shows considerably more
structure, as well as additional variation due to the curvature of the
hard surface.  The symmetric correlation function is nonzero at
locations where spheres may touch, which for a convex hard surface
means that $g_\sigma^S$ may be nonzero in the volume in which hard
spheres are excluded.  In every configuration studied, the agreement
between the theoretical predictions and the Monte Carlo simulation in
each case is very poor in the region where there should be no contacts
at all.  Because $n_0$ is comparable to its bulk value in this region,
this means that these functionals predict a significant number of
contacts in the region where there should be none.  The correlation
function of Yu and Wu~\cite{yu2002fmt-dft-inhomogeneous-associating}
and ours described in Section~\ref{sec:g-S} give similar results, with
slightly larger errors in our prediction.

\subsection{High density}

\begin{figure}
  \includegraphics[width=\columnwidth]{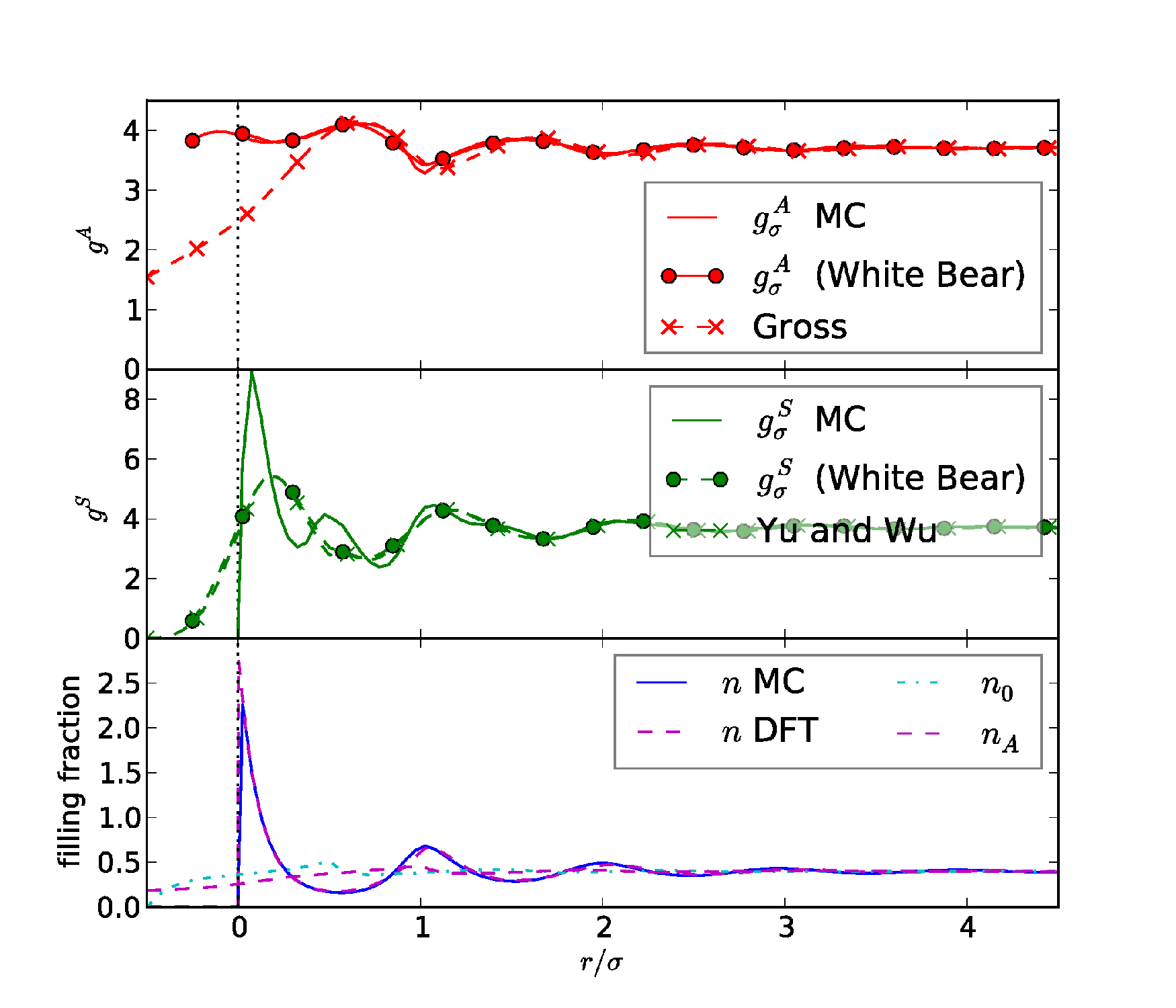}
  \caption{Density and correlation function at a hard wall with bulk
    filling fraction of 0.4.}
  \label{fig:walls-40}
\end{figure}

\begin{figure}
  \includegraphics[width=\columnwidth]{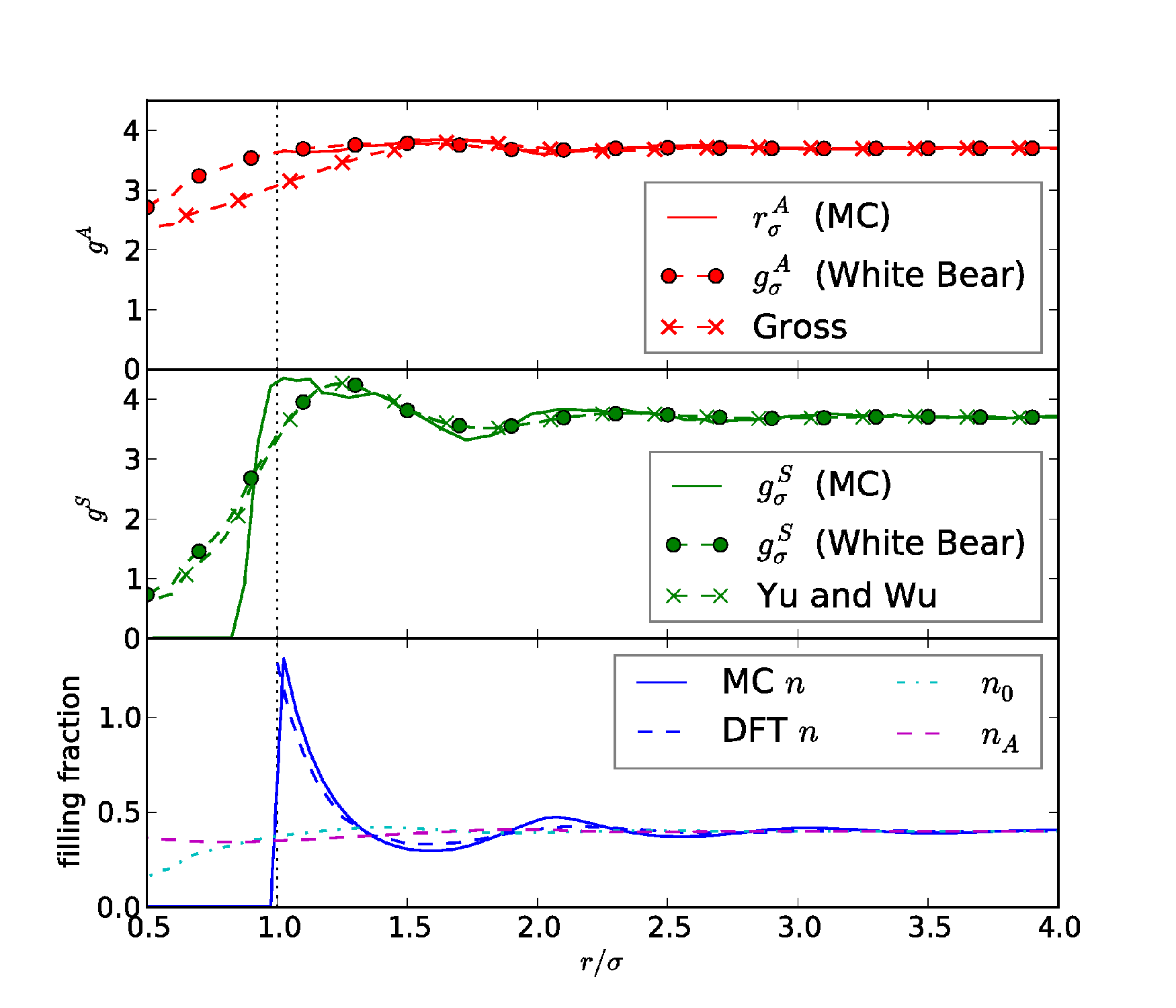}
  \caption{Density and correlation function around a hard sphere with
    diameter the same as those in the fluid, with a bulk filling
    fraction of 0.4.}
  \label{fig:inner-4-40}
\end{figure}

\begin{figure}
  \includegraphics[width=\columnwidth]{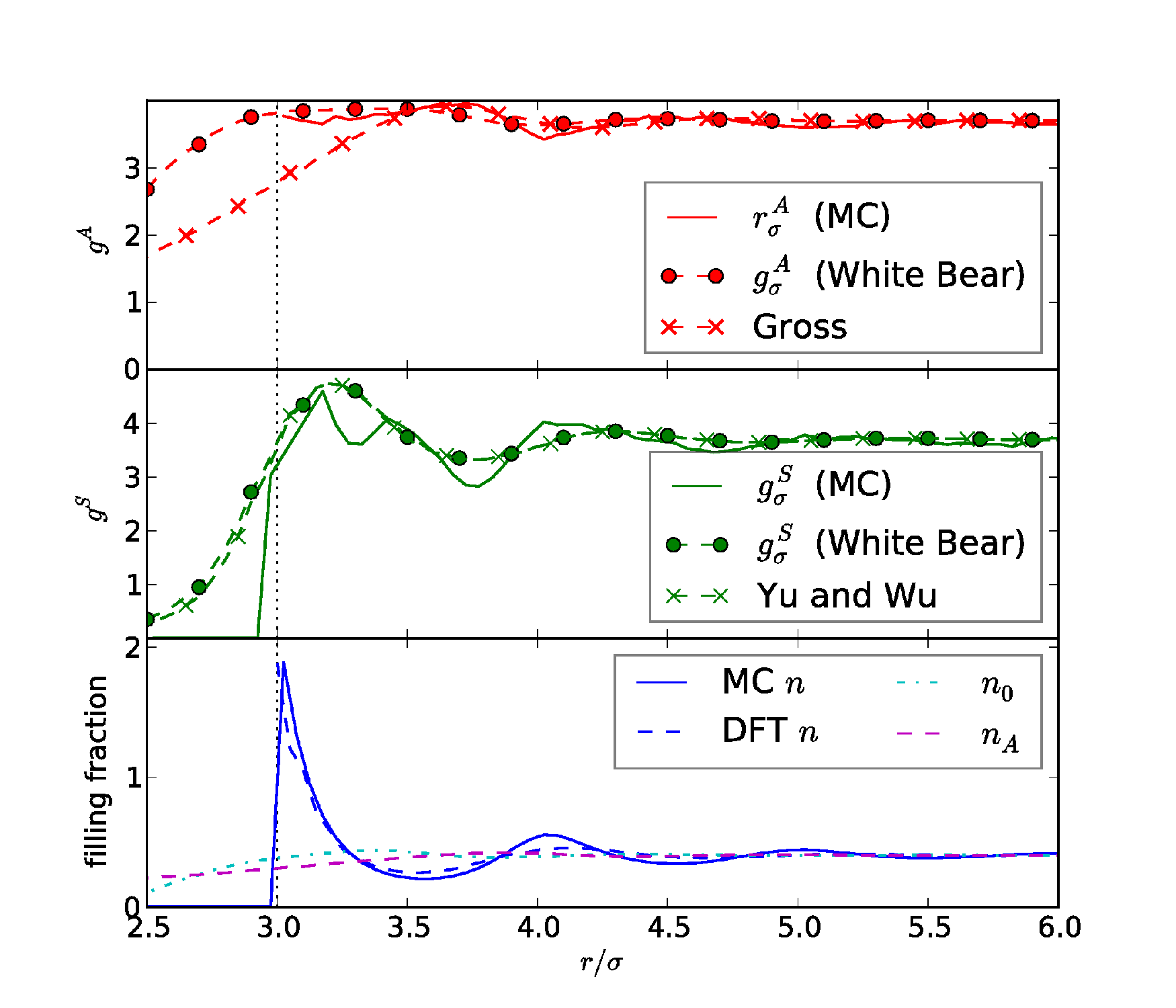}
  \caption{Density and correlation function around a hard sphere with
    an excluded diameter of $6\sigma$, with a bulk filling fraction of
    0.4.}
  \label{fig:inner-12-40}
\end{figure}

\begin{figure}
  \includegraphics[width=\columnwidth]{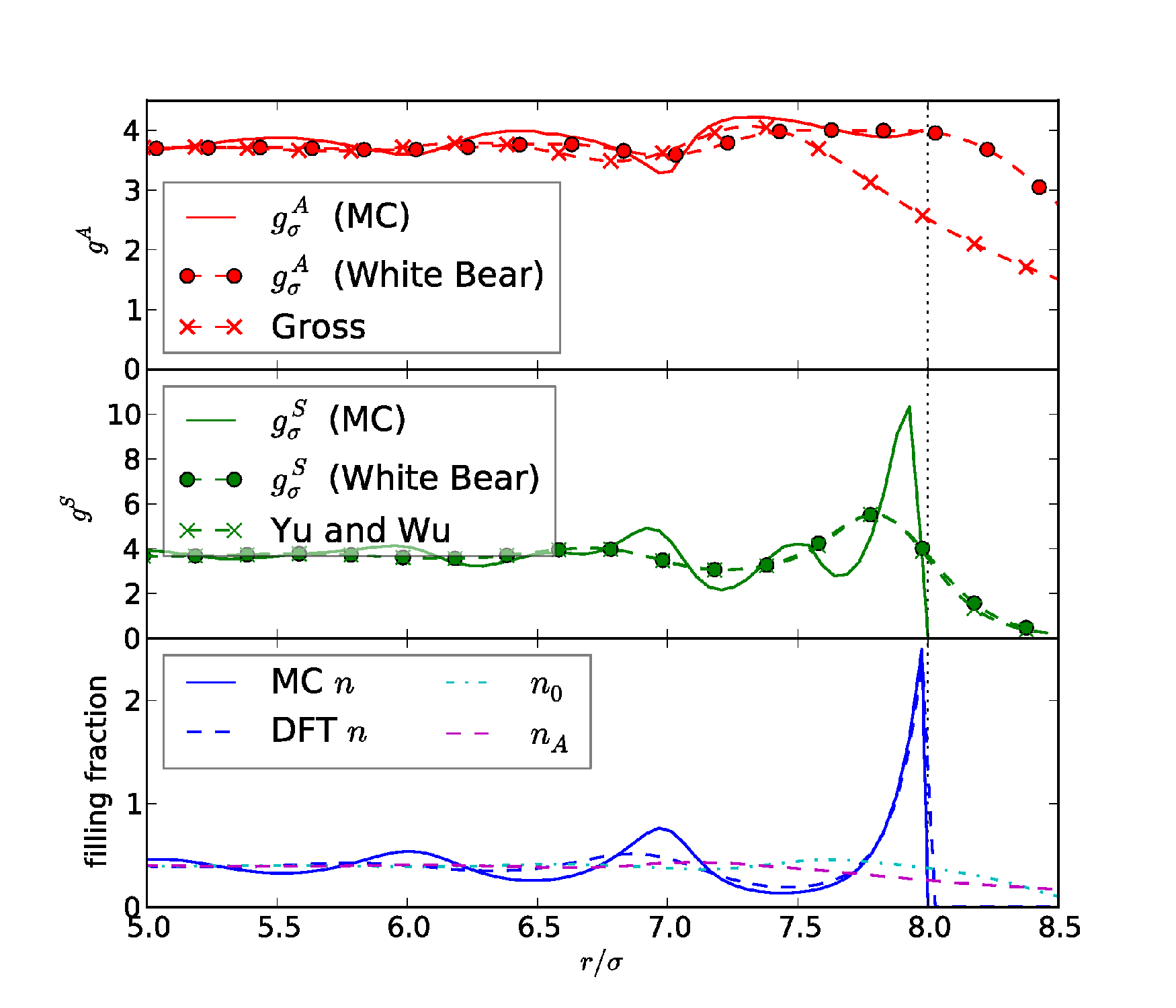}
  \caption{Density and correlation functions near the surface of a
    spherical cavity with diameter $16\sigma$ at a bulk filling
    fraction of 0.4.}
  \label{fig:outer-40}
\end{figure}

At a higher density corresponding to a filling fraction of 0.4,
correlations are much stronger, with the bulk contact value of the
correlation function of 3.7, as seen in
Figures~\ref{fig:walls-40}-\ref{fig:outer-40}.  This results in larger
oscillations in the density at the hard surfaces, and correspondingly
more interesting behavior in the correlation function near the
interface.  The density predicted by the White Bear functional agrees
reasonably well with the simulation results, although not so well as
it did at lower density.  The discrepancies are largest in the case of
the spherical cavity, in which the DFT considerably underestimates the
range of the density oscillations.

The asymmetric version of the correlation function once again displays
relatively smooth behavior with a few small
oscillations near the interface, and a somewhat elevated value within
a diameter of the hard surface, with the magnitude of this elevation
somewhat different in each configuration.  As was the case at low
density, our correlation function $g_\sigma^A$ matches very closely
the Monte Carlo data, reproducing quite well the structure near the
interface in each configuration, although in the spherical cavity
(Figure~\ref{fig:outer-40}), there is a small, but significant
discrepancy, comparable to the discrepancy found in the density
itself.  In each case, the correlation of Gross dramatically
underestimates the correlation at the interface, at one extreme by
40\% in the case of the spherical cavity
(Figure~\ref{fig:outer-40}), and at the other extreme by 15\%
in the test-particle scenario (Figure~\ref{fig:inner-4-40}).

The symmetrically averaged correlation function shows considerably
more structure near the interface at high density, and this structure
varies considerably depending on the curvature of the hard surface.
In each case, this structure is not reflected in the theoretical
predictions, neither that of this paper, nor that of Yu and
Wu~\cite{yu2002fmt-dft-inhomogeneous-associating}.  As was the case at
low density, both functionals give significant and finite values in
the region in which there are no contacts, but at high density they
also miss the large oscillations that are present near the flat wall
and the concave surface (Figures~\ref{fig:walls-40}
and~\ref{fig:outer-40}).  As was the case at low density, the
functional of Yu and Wu~\cite{yu2002fmt-dft-inhomogeneous-associating}
gives slightly better agreement with the simulation results than that
which we derive in Section~\ref{sec:g-S}.

\section{Conclusion}
We investigated several approximations to the contact value of the
correlation function for inhomogeneous fluid distributions
corresponding to flat, concave, and convex walls.  We
defined and simulated two averages of the correlation function, an
asymmetric $A$ average centered at the location of one of the two
spheres that is in contact, and a symmetric $S$ average centered at
the point of contact of touching spheres.  For each average, we
derived a functional form from FMT, and also found an approximation
that has been used in the literature.  When compared with essentially
exact Monte Carlo simulations, the $A$ correlation function derived
from Fundamental Measure Theory in Section~\ref{sec:g-A} gives
excellent results for each surface, at both high density and low
density.  The other three approximations that we studied all showed
significant and systematic deviations under some circumstances.  Thus,
we recommend that creators of SAFT-based classical density functionals
consider using the $g_\sigma^A$ functional defined in
Section~\ref{sec:g-A}.

\appendix



\section*{Appendix}

The expression for the asymmetric correlation function
$g_\sigma^A(\rr)$ (Equation~\ref{eq:g-A-exact}) involves the
functional derivative $\frac{\delta A_{HS}}{\delta
  \sigma(\mathbf{r})}$.  In this appendix we will explain how this
derivative is evaluated.  We begin by applying the chain rule in the
following way:
  \begin{align}
    \frac{\delta A_{HS}}{\delta \sigma(\mathbf{r})} &=
    \int \left(
    \sum_\alpha
    \frac{\delta A_{HS}}{\delta n_\alpha(\mathbf{r}')}
    \frac{\delta n_\alpha(\mathbf{r}')}{\delta \sigma(\mathbf{r})}
    \right) d\mathbf{r}'
  \end{align}
This expression requires us to evaluate $\frac{\delta A_{HS}}{\delta
  n_\alpha(\mathbf{r}')}$ and $\frac{\delta
  n_\alpha(\mathbf{r}')}{\delta \sigma(\mathbf{r})}$.  The former is
straightforward, given Equations~\ref{eq:Phi1}-\ref{eq:Phi3}, and we
will write no more about it.  The functional derivatives of the
fundamental measures, however, require a bit more subtlety, and we
will address them here.

We begin with the derivative of $n_3$, the filling fraction, which we
will discuss in somewhat more detail than the remainder, which are
similar in nature.  Because the diameter $\sigma(\rr)$ is the diameter
of a sphere \emph{at position~$\rr$}, we write the fundamental measure
$n_3(\rr')$ as
\begin{align}
  n_3(\rr') &= \int n(\rr'') \Theta\left(\frac{\sigma(\rr'')}{2}
  -\left|\rr' - \rr''\right|\right)
  d\rr''
\end{align}
where we note that $\sigma(\rr'')$ and $n(\rr'')$ are the diameter and
density, respectively, of spheres centered at position~$\rr''$.  Thus the
derivative with respect to the diameter of spheres at position
$\rr$ is
\begin{align}
  \frac{\delta n_3(\rr')}{\delta \sigma(\rr)} &= \frac 12 \int n
  (\rr'') \delta\left(\frac{\sigma(\rr'')}{2} -\left|\rr' - \rr''\right|\right)
  \delta(\rr-\rr'') d\rr'' \\ &= n (\rr) \delta(\sigma(\rr)/2
  -\left|\rr' - \rr\right|)
\end{align}
This pattern will hold for each fundamental measure: because we are
seeking the change in free energy when spheres at point~$\rr$ are
expanded, the integral over density is eliminated.  To compute the
correlation funtion $g_\sigma^A$, we convolve this delta function with
the product of the density and a local derivative of $\Phi(\rr)$:
\begin{align}
  \frac{\delta A_{HS}}{\delta \sigma(\rr)} &= \int \frac{\partial \Phi(\rr')}{\partial
    n_3(\rr')}n(\rr') \delta(\sigma/2-|\rr'-\rr|)d\rr'
  + \cdots
\end{align}
As we shall see, there are only four convolution kernels, leading to
four additional convolutions beyond those required for FMT.

The functional derivative of $n_2$ introduces our second convolution
kernel, which is a derivative of the delta function.
\begin{align}
  \frac{\delta n_2(\rr')}{\delta \sigma(\rr)} &= \frac 12 n(\rr) \delta'(\sigma(\rr)/2 -\left|\rr' - \rr\right|)
\end{align}
The derivatives of the remaining scalar densities $n_1$ and $n_0$ reduce to
sums of the terms above:
\begin{multline}
  \frac{\delta n_1(\rr')}{\delta \sigma(\rr)}
  = \frac{n(\rr)}{4\pi
    \sigma(\rr)}\delta'(\sigma(\rr)/2 -\left|\rr' - \rr\right|) \\
  -
  \frac{n(\rr)}{2\pi
    \sigma(\rr)^2}\delta(\sigma(\rr)/2 -\left|\rr' - \rr\right|)
\end{multline}
and
\begin{multline}
  \frac{\delta n_0(\rr')}{\delta \sigma(\rr)}
  = \frac{n(\rr)}{2\pi
    \sigma(\rr)^2}\delta'(\sigma(\rr)/2 -\left|\rr' - \rr\right|)
  \\-
  2 \frac{n(\rr)}{\pi
    \sigma(\rr)^3}\delta(\sigma(\rr)/2 -\left|\rr' - \rr\right|)
\end{multline}

The vector-weighted densities $\mathbf{n}_{V1}$ and $\mathbf{n}_{V2}$
give terms analogous to those of $n_1$ and $n_2$:
\begin{multline}
  \frac{\delta \mathbf{n}_{V2}(\rr')}{\delta \sigma(\rr)} = -\frac 12 n(\rr) \delta'(\sigma(\rr)/2 -\left|\rr' - \rr\right|)
    \frac{\rr-\rr'}{|\rr-\rr'|}
\end{multline}
\begin{multline}
  \frac{\delta \mathbf{n}_{V1}(\rr')}{\delta \sigma(\rr)}
  = -\frac{n(\rr)}{4\pi
    \sigma(\rr)}\delta'(\sigma(\rr)/2 -\left|\rr' - \rr\right|) \frac{\rr-\rr'}{|\rr-\rr'|}
  \\ +
  \frac{n(\rr)}{2\pi
    \sigma(\rr)^2}\delta(\sigma(\rr)/2 -\left|\rr' - \rr\right|) \frac{\rr-\rr'}{|\rr-\rr'|}
\end{multline}
Thus there are four convolution kernels used in computing $g_\sigma^A$:
one scalar and one vector delta function, and one scalar and one
vector derivative of the delta function.

\bibliography{paper}

\end{document}